\documentclass[aps,prb,twocolumn,showpacs,superscriptaddress]{revtex4}

\usepackage{graphicx}
\usepackage{amsmath}
\usepackage{amssymb}

\begin{document}

\title{Effect of the pseudogap on the infrared response in cuprate superconductors}


\author{Ling Qin}

\affiliation{Department of Physics, Beijing Normal University, Beijing 100875, China}

\author{Jihong Qin}

\affiliation{Department of Physics, University of Science and Technology Beijing, Beijing 100083, China}

\author{Shiping Feng}

\affiliation{Department of Physics, Beijing Normal University, Beijing 100875, China}

\begin{abstract}
One of the most essential aspects of cuprate superconductors is a large pseudogap coexisting with a superconducting gap, then some anomalous properties can be understood in terms of the formation of the pseudogap. Within the kinetic energy driven superconducting mechanism, the effect of the pseudogap on the infrared response of cuprate superconductors in the superconducting-state is studied. By considering the interplay between the superconducting gap and pseudogap, the electron current-current correlation function is evaluated based on the linear response approach and it then is employed to calculate finite-frequency conductivity. It is shown that in the underdoped and optimally doped regimes, the transfer of the part of the low-energy spectral weight of the conductivity spectrum to the higher energy region to form a midinfrared band is intrinsically associated with the presence of the pseudogap.
\end{abstract}


\maketitle

The parent compounds of cuprate superconductors are identified as Mott insulators \cite{Kastner98}, in which the lack of conduction arises from anomalously strong electron-electron repulsion. Superconductivity then is obtained by adding charge carriers to insulating parent compounds with the superconducting (SC) transition temperature $T_{\rm c}$ takes a domelike shape with the underdoped and overdoped regimes on each side of the optimal doping, where $T_{\rm c}$ reaches its maximum \cite{Tallon95}. However, a number of experimental probes \cite{Deutscher05,Devereaux07,Hufner08,Puchkov96,Timusk99,Basov05} show that below a characteristic temperature $T^{*}$, which can be well above $T_{\rm c}$ in the underdoped and optimally doped regimes, the physical response of cuprate superconductors can be interpreted in terms of the formation of a normal-state pseudogap by which it means a suppression of the spectral weight of the low-energy excitation spectrum. In particular, this normal-state pseudogap crossover temperature $T^{*}$ decreases with increasing doping in the underdoped regime and since $T_{\rm c}$ rises with doping, then $T^{*}$ seems to merge with $T_{\rm c}$ in the overdoped regime, eventually disappearing together with superconductivity at the end of the SC dome \cite{Deutscher05,Devereaux07,Hufner08,Puchkov96,Timusk99,Basov05}. After intensive investigations over more than two decades, it has become clear that the normal-state pseudogap is thought to be key to understanding the mechanism of superconductivity in cuprate superconductors.

The infrared measurements of the reflectance are an ideal way to study the low-energy excitations of cuprate superconductors \cite{Puchkov96,Timusk99,Basov05,Puchkov96a,Basov96,Lee05,Hwang07,Mirzaei13}. In particular, it has been shown in terms of the Kramers-Kronig analysis of the reflectance that the SC-state conductivity is rather universal within the whole cuprate superconductors \cite{Puchkov96,Timusk99,Basov05,Puchkov96a,Basov96,Lee05,Hwang07,Mirzaei13}, where a key feature is the two-component conductivity: a non-Drude-like narrow band centered around energy $\omega\sim 0$ followed by a broadband centered in the midinfrared region in the underdoped and optimally doped regimes. In particular, this two-component conductivity extends to the normal-state pseudogap boundary in the phase diagram at $T^{*}$ \cite{Lee05,Hwang07,Mirzaei13}. However, this anomalous structure of the SC-state conductivity spectrum can not be described by a simple Bardeen-Cooper-Schrieffer (BCS) formalism \cite{Schrieffer83}. There is mounting evidence that the unusual feature of the SC-state conductivity spectrum in the underdoped and optimally doped regimes is dominated by the normal-state pseudogap  \cite{Puchkov96,Timusk99,Basov05,Puchkov96a,Basov96,Lee05,Hwang07,Mirzaei13}. In our recent study \cite{Qin14}, the doping and temperature dependence of the optical conductivity of cuprate superconductors in the normal-state has been discussed by considering the effect of the normal-state pseudogap on the optical response, and then the main feature of the optical measurements on cuprate superconductors in the normal-state is qualitatively reproduced. As a complement of our previous analysis of the optical conductivity in the normal-state, we in this paper discuss the conductivity of cuprate superconductors in the SC-state based on the kinetic energy driven SC  mechanism \cite{Feng0306} by considering the interplay between the SC gap and normal-state pseudogap \cite{Feng12}. In particular, we show that the part of the low-energy spectral weight of the conductivity spectrum in the underdoped and optimally doped regimes is transferred to the higher energy region to form the unusual midinfrared band, however, the onset of the region to which the spectral weight is transferred is always close to the normal-state pseudogap.

The single common element of cuprate superconductors is two-dimensional CuO$_{2}$ planes \cite{Kastner98}. As originally emphasized by Anderson \cite{Anderson87}, the essential physics of the doped CuO$_{2}$ plane is properly captured by the $t$-$J$ model acting on the space with no doubly occupied sites. This $t$-$J$ model consists of two parts, the kinetic energy part includes the nearest-neighbor (NN) hopping term $t$ and next NN hopping term $t'$, while the magnetic energy part is described by a Heisenberg term with the NN spin-spin antiferromagnetic (AF) exchange $J$. The high complexity in the $t$-$J$ model comes mainly from the local constraint of no double electron occupancy, i.e., $\sum_{\sigma}C^{\dagger}_{l\sigma}C_{l\sigma}\leq 1$, which can be treated properly within the charge-spin separation (CSS) fermion-spin theory \cite{Feng0494}, where the constrained electron operators $C_{l\uparrow}$ and $C_{l\downarrow}$ are decoupled as $C_{l\uparrow}=h^{\dagger}_{l\uparrow}S^{-}_{l}$ and $C_{l\downarrow}=h^{\dagger}_{l\downarrow}S^{+}_{l}$, respectively, with the spinful fermion operator $h_{l\sigma}=e^{-i\Phi_{l\sigma}}h_{l}$ that keeps track of the charge degree of freedom together with some effects of spin configuration rearrangements due to the presence of the doped hole itself (charge carrier), while the spin operator $S_{l}$ represents the spin degree of freedom, then the local constraint of no double electron occupancy is satisfied in analytical calculations. In this CSS fermion-spin representation, the magnetic energy term in the $t$-$J$ model is only to form an adequate spin configuration, while the kinetic energy is transferred as the interaction between charge carriers and spins, and therefore dominates the essential physics in cuprate superconductors.

Based on the $t$-$J$ model in the CSS fermion-spin representation, we \cite{Feng0306} have developed a kinetic energy driven SC mechanism. In particular, the interplay between the normal-state pseudogap state and superconductivity has been discussed \cite{Feng12} within the framework of the kinetic energy driven SC mechanism, where the interaction between charge carriers and spins directly from the kinetic energy in the $t$-$J$ model by exchanging spin excitations induces the SC-state in the particle-particle channel and the pseudogap state in the particle-hole channel, then there is a coexistence of the SC gap and pseudogap in the whole SC dome. This pseudogap antagonizes superconductivity, and then $T_{\rm c}$ is suppressed to low temperatures. Following our previous discussions \cite{Feng12}, the full charge carrier diagonal and off-diagonal Green's functions of the $t$-$J$ model in the SC-state can be evaluated as,
\begin{widetext}
\begin{subequations}\label{hole-Green-function}
\begin{eqnarray}
g({\bf k},\omega)&=&{1\over \omega-\xi_{\bf k}-\Sigma^{({\rm h})}_{1}({\bf k},\omega)-\bar{\Delta}^{2}_{\rm h}({\bf k})/[\omega+\xi_{\bf k}
+\Sigma^{({\rm h})}_{1}({\bf k}, -\omega)]},\label{hole-diagonal-Green's-function}\\
\Gamma^{\dagger}({\bf k},\omega)&=&-{\bar{\Delta}_{\rm h}({\bf k})\over [\omega-\xi_{\bf k}-\Sigma^{({\rm h})}_{1}({\bf k},\omega)][\omega+\xi_{\bf k}
+\Sigma^{({\rm h})}_{1}({\bf k},-\omega)]-\bar{\Delta}^{2}_{\rm h}({\bf k})},\label{hole-off-diagonal-Green's-function}
\end{eqnarray}
\end{subequations}
\end{widetext}
with the self-energy in the particle-hole channel,
\begin{eqnarray}\label{self-energy}
\Sigma^{({\rm h})}_{1}({\bf k},\omega)\approx {[2\bar{\Delta}_{\rm pg}({\bf k})]^{2}\over \omega+M_{\bf k}},
\end{eqnarray}
where we use the same notations as in Ref. \cite{Feng12}, and in particular, the mean-field charge carrier spectra $\xi_{\bf k}$, $M_{\bf k}$, the effective d-wave charge carrier pair gap $\bar{\Delta}_{\rm h}({\bf k})$, and the effective normal-state pseudogap $\bar{\Delta}_{\rm pg}({\bf k})$ have been given explicitly in Ref. \cite{Feng12}. This kinetic energy driven SC mechanism \cite{Feng0306,Feng12} also indicates that the strong electron correlation favors superconductivity, since the main ingredient is identified into a charge carrier pairing mechanism not involving the phonon, the external degree of freedom, but the internal spin degree of freedom of electron.

Through the standard linear response theory \cite{Mahan81}, the finite-frequency conductivity of cuprate superconductors can be expressed as, $\sigma(\omega)=-{\rm Im} \Pi(\omega) /\omega$, where $\Pi(\tau-\tau')=-\langle T_{\tau}{\bf j}(\tau)\cdot {\bf j}(\tau')\rangle$ is the electron current-current correlation function. This electron current operator ${\bf j}(\tau)$ is obtained in terms of the electron polarization operator ${\bf P}$, which is a summation over all the particles and their positions \cite{Mahan81}, and can be given as ${\bf P}= -e\sum\limits_{l\sigma}{\bf R}_{l}C^{\dagger}_{l\sigma}C_{l\sigma}$. The external magnetic field can be coupled to the electrons, which are now represented by $C_{l\uparrow}=h^{\dagger}_{l\uparrow}S^{-}_{l}$ and $C_{l\downarrow}= h^{\dagger}_{l\downarrow}S^{+}_{l}$ in the CSS fermion-spin representation \cite{Feng0494}. In this CSS fermion-spin representation, the electron current operator can be decoupled as the charge carrier and spin parts, respectively. In particular, we \cite{Qin14} have shown that there is no direct contribution to the electron current operator from the electron spin part, and then the main contribution for the electron current operator comes from the charge carriers (then the electron charge), however, the strong interplay between the charge carriers and spins has been considered through the spin's order parameters entering in the charge carrier part of the contribution to the current-current correlation. In this case, the finite-frequency conductivity of cuprate superconductors in the SC-state can be evaluated as,
\begin{widetext}
\begin{eqnarray}\label{conductivity}
\sigma(\omega) &=& \left ({Ze\over \hbar}\right )^{2}{1\over N}\sum_{\bf k}\gamma^{2}_{{\rm s}{\bf k}}\int^{\infty}_{-\infty}{{\rm d}\omega'\over 2\pi}[A_{g}({\bf k},\omega+\omega') A_{g}({\bf k},\omega')+A_{\Gamma}({\bf k},\omega+\omega')A_{\Gamma}({\bf k},\omega')]{n_{\rm F}(\omega')-n_{\rm F}(\omega+\omega')\over\omega},
\end{eqnarray}
\end{widetext}
width $n_{\rm F}(\omega)$ is the fermion distribution function, while the spectral functions $A_{g}({\bf k},\omega)$ and $A_{\Gamma}({\bf k},\omega)$ are obtained in terms of the charge carrier diagonal and off-diagonal Green's functions in Eq. (\ref{hole-Green-function}) as $A_{g}({\bf k},\omega)=-2{\rm Im}g({\bf k},\omega)$ and $A_{\Gamma}({\bf k},\omega) =-2{\rm Im}\Gamma^{\dagger}({\bf k},\omega)$, respectively.

\begin{figure}[h!]
\center\includegraphics[scale=0.32]{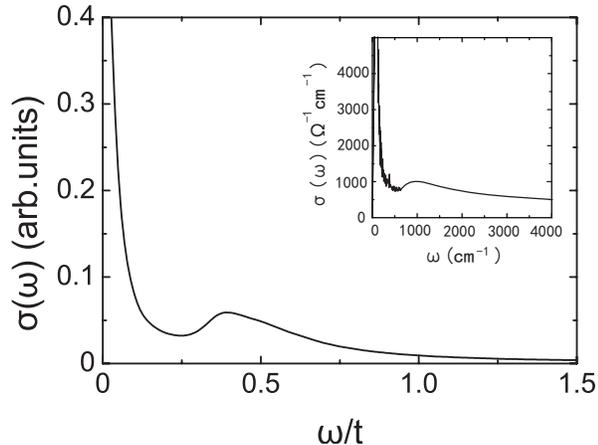}
\caption{The conductivity as a function of energy in $\delta=0.09$ with $T=0.002J$ for $t/J=2.5$ and $t'/t=0.3$. Inset: the corresponding experimental data of the underdoped Bi$_{2}$Sr$_{2}$CaCu$_{2}$O$_{8+\delta}$ taken from Ref. \onlinecite{Hwang07}. \label{fig1}}
\end{figure}

In Fig. \ref{fig1}, we plot the results of the SC-state conductivity (\ref{conductivity}) as a function of energy in the underdoping $\delta=0.09$ for parameters $t/J=2.5$ and $t'/t=0.3$ with temperature $T=0.002J$, where the charge $e$ has been set as the unit. For comparison, the corresponding experimental result \cite{Hwang07} of the underdoped Bi$_{2}$Sr$_{2}$CaCu$_{2}$O$_{8+\delta}$ is also plotted in Fig. \ref{fig1} (inset). Obviously, the experimental data of cuprate superconductors obtained from the conductivity measurements are qualitatively reproduced \cite{Puchkov96,Timusk99,Basov05,Puchkov96a,Basov96,Lee05,Hwang07,Mirzaei13}. At the low temperatures, the SC-state conductivity $\sigma(\omega)$ extends over a broad energy range, and consists of a low-energy component and a higher energy band separated by a gap at $\omega\sim 0.2t$. This higher energy band, corresponding to the "midinfrared band", shows a broad peak at $\omega\sim 0.38t$, and persists in the normal-state pseudogap phase \cite{Qin14}. The low-energy component forms a sharp peak at $\omega\sim 0$, however, it deviates strongly from the Drude behavior, and has a power-law decay form as $\sigma(\omega)\rightarrow 1/\omega$ (the non-Drude formula). In comparison with our previous results of the optical conductivity in the normal-state \cite{Qin14}, we find that the spectral weight of the low-energy component has been further suppressed by the SC gap, however, there is no depletion of the spectral weight of the higher energy midinfrared band, which is consistent with the experimental observation on cuprate superconductors \cite{Lee05}. Moreover, as in the case of the normal-state \cite{Qin14}, the onset of the region to which the spectral weight is transferred is close to the normal-state pseudogap $\bar{\Delta}_{\rm pg}$, reflecting a fact that due to the presence of the normal-state pseudogap in cuprate superconductors, the part of the low-energy spectral weight of the SC-state conductivity spectrum in the underdoped regime is transferred to the higher energy region to form the unusual midinfrared band. In the other words, the appearance of the higher energy midinfrared band is closely related to the effect of the normal-state pseudogap on the infrared response in cuprate superconductors \cite{Lee05,Hwang07,Mirzaei13,Yu08,Hwang08}.

\begin{figure}[h!]
\center\includegraphics[scale=0.32]{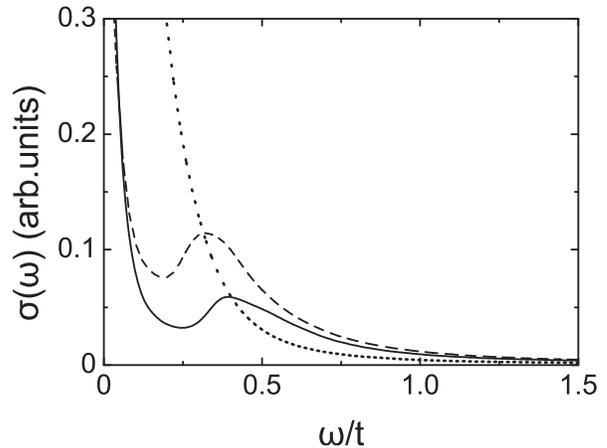}
\caption{The conductivity as a function of energy in $\delta=0.09$ (solid line), $\delta=0.15$ (dashed line), and $\delta=0.25$ (dotted line) with $T=0.002J$ for $t/J=2.5$ and $t'/t=0.3$.\label{fig2}}
\end{figure}

The spectral weight is proportional to the area under the conductivity curve in Fig. \ref{fig1}. However, the weight and position of the midinfrared band are strongly doping dependent. To see this point clearly, we have made a series of calculations for the SC-state conductivity (\ref{conductivity}) in the whole doping range from the underdoped to heavily overdoped, and the results of $\sigma(\omega)$ as a function of energy in the underdoping $\delta=0.09$ (solid line), the optimal doping $\delta=0.15$ (dashed line), and the heavy overdoping $\delta=0.25$ (dotted line) for $t/J=2.5$ and $t'/t=0.3$ with $T=0.002J$ are plotted in Fig. \ref{fig2}. It is shown clearly that as the charge carrier doping increases, the overall conductivity increases. Within the framework of the kinetic energy driven SC mechanism \cite{Feng0306}, the magnitude of the normal-state pseudogap $\bar{\Delta}_{\rm pg}$ is much larger than that of the SC gap in the underdoped regime, then it smoothly decreases upon increasing doping, eventually disappearing together with superconductivity at the end of the SC dome \cite{Feng12}. In corresponding to this evolution of the normal-state pseudogap with doping, the positions of the gap and midinfrared peak gradually shift to the lower energies with increasing doping, i.e., there is a tendency that with increasing doping, the magnitude of the gap decreases, then the higher energy midinfrared band moves towards to the low-energy non-Drude band. This tendency is particularly obvious in the overdoped regime. In particular, in the heavily overdoped regime, the normal-state pseudogap is negligible \cite{Feng12}, i.e., $\bar{\Delta}_{\rm pg}\approx 0$, then the full charge carrier diagonal and off-diagonal Green's functions in Eq. (\ref{hole-Green-function}) are reduced as a simple d-wave BCS formalism. It is thus similar to the conventional superconductors \cite{Schrieffer83} except the d-wave symmetry. In this case, the low-energy non-Drude peak incorporates with the midinfrared band, and then the low-energy Drude type behavior  ($\sigma(\omega)\rightarrow 1/\omega^{2}$) of the conductivity recovers in the heavily overdoped regime, in qualitative agreement with the corresponding experimental data of cuprate superconductors \cite{Puchkov96,Timusk99,Basov05,Puchkov96a,Basov96,Lee05,Hwang07,Mirzaei13}.

\begin{figure}[h!]
\center\includegraphics[scale=0.32]{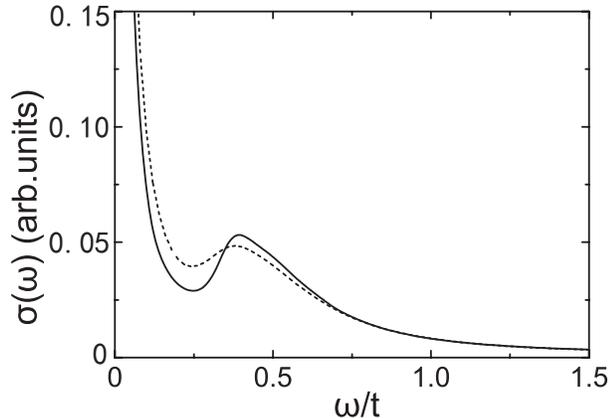}
\caption{The conductivity as a function of energy in $\delta=0.09$ with $T=0.002J$ (solid line) and $T=0.146J$ (dashed line) for $t/J=2.5$ and $t'/t=0.3$.\label{fig3}}
\end{figure}

Within the framework of the kinetic energy driven SC mechanism, the calculated \cite{Feng12} $T_{\rm c}\sim 0.06J$ and $T^{*}\sim 0.19J$ at doping $\delta=0.09$. The low-energy non-Drude component of the conductivity spectrum in Fig. \ref{fig1} broadens as temperature increases from the temperature $T\ll T_{\rm c}$ to $T>T_{\rm c}$ and continues to grow even just above the normal-state pseudogap crossover temperature $T^{*}$. However, this broadness of the low-energy component of the conductivity spectrum is accompanied by a decrease of the weight of the higher energy midinfrared band, and then the low-energy Drude type behavior of the conductivity recovers at the temperature above $T^{*}$. To see this point clearly, in Fig. \ref{fig3}, we plot $\sigma(\omega)$ as a function of energy with $T=0.002J$ (solid line) in the SC-state and $T=0.146J$ (dashed line) in the normal-state pseudogap phase in doping $\delta=0.09$ for $t/J=2.5$ and $t'/t=0.3$. In particular, the results in Fig. \ref{fig3} confirm again that in spite of the conductivity in the SC-state or the normal-state pseudogap phase \cite{Qin14}, the onset of the region to which the spectral weight is transferred is always close to the normal-state pseudogap $\bar{\Delta}_{\rm pg}$ \cite{Lee05,Hwang07,Mirzaei13,Yu08,Hwang08}, also in qualitative agreement with the corresponding experimental data of cuprate superconductors \cite{Puchkov96,Timusk99,Basov05,Puchkov96a,Basov96,Lee05,Hwang07,Mirzaei13}.

Although the separate normal-state pseudogap and SC gap are underlying the redistribution of the spectral weight in the SC-state conductivity spectrum, the essential physics of the higher energy midinfrared band in cuprate superconductors in the SC-state is the same as in the case of the normal-state, and can be attributed to the emergence of the normal-state pseudogap. This follows a fact that in Eq. (\ref{conductivity}), there are two parts of the charge carrier quasiparticle contribution to the redistribution of the spectral weight in the SC-state conductivity: the contribution from the first term of the right-hand side in Eq. (\ref{conductivity}) comes from the spectral function obtained in terms of the charge carrier diagonal Green's function (\ref{hole-diagonal-Green's-function}), and therefore is closely associated with the normal-state pseudogap $\bar{\Delta}_{\rm pg}$ in the particle-hole channel, while the additional contribution from the second term of the right-hand side in Eq. (\ref{conductivity}) originates from the spectral function obtained in terms of the charge carrier off-diagonal Green's function (\ref{hole-off-diagonal-Green's-function}), and is closely related to the charge carrier pair gap $\bar{\Delta}_{\rm h}$ in the particle-particle channel. However, since $\bar{\Delta}_{\rm h}\ll\bar{\Delta}_{\rm pg}$ in the underdoped and optimally doped regimes \cite{Feng12}, the SC gap only suppresses the spectral weight of the low-energy component, while the normal-state pseudogap related shift of the spectral weight from the low-energy to the higher energy midinfrared band in the SC-state conductivity spectrum becomes arrested. Since both the normal-state pseudogap state and superconductivity in cuprate superconductors are the result of the strong electron correlation within the framework of the kinetic energy driven SC mechanism \cite{Feng12}, the transfer of the part of the low-energy spectral weight of the conductivity spectrum in the underdoped and optimally doped regimes to the higher energy region to form the unusual midinfrared band is a natural consequence of the strongly correlated nature in cuprate superconductors. In particular, this strong electron correlation induces a shift of the spectral weight from the low-energy to the higher energy midinfrared band in the conductivity spectrum has been confirmed by the early numerical simulations based on the $t$-$J$ model in the normal-state \cite{Stephan90,Dagotto94} and the SC-state \cite{Haule07}. In an ordinary metal, the shape of the conductivity is normally well accounted for by the low-energy Drude formula ($\sigma(\omega)\rightarrow 1/\omega^{2}$) that describes the charge carrier contribution to the conductivity, then when the temperature $T<T_{\rm c}$, the spectral weight of the condensate in the SC-state comes from low energies \cite{Schrieffer83}. However, in cuprate superconductors, since the higher energy midinfrared band is taken from the low-energy band, and the onset of the region to which the spectral weight is transferred is close to the normal-state pseudogap, so that the large normal-state pseudogap in the underdoped and optimally doped regimes heavily reduces the fraction of the charge carriers that condense in the SC-state \cite{Homes04}.

In conclusion, we have shown very clearly in this paper that if the interplay between the SC gap and normal-state pseudogap is taken into account in the framework of the kinetic energy driven SC mechanism, the SC-state conductivity of the $t$-$J$ model calculated based on the linear response approach per se can correctly reproduce the main feature found in the infrared response measurements on cuprate superconductor in the SC-state. Our results also show that the transfer of the part of the low-energy spectral weight in the SC-state conductivity in the underdoped and optimally doped regimes to the higher energy region to form the unusual midinfrared band can be attributed to the effect of the normal-state pseudogap on the infrared response in cuprate superconductors.

\acknowledgments

The authors would like to thank Dr. Huaisong Zhao for the helpful discussions. LQ and SF are supported by the National Natural Science Foundation of China under Grant No. 11274044, and the funds from the Ministry of Science and Technology of China under Grant Nos. 2011CB921700 and 2012CB821403, and JQ is supported by the Beijing Higher Education Young Elite Teacher Project under Grant No. 0389.

\end{document}